\documentclass[%
 reprint,
 amsmath,amssymb,
 aps,
]{revtex4-2}
\usepackage[utf8]{inputenc}
\usepackage{graphicx}
\usepackage[colorlinks,linkcolor=blue,citecolor=blue]{hyperref}
\usepackage{breakurl}
\usepackage{array}
\usepackage{color}
\usepackage{amsmath}
\usepackage{amsxtra}
\usepackage{amstext}
\usepackage{amssymb}
\usepackage{physics}
\usepackage{latexsym}
\usepackage{float}

\graphicspath{./Figures/}

\begin{document} 

\title{Ultrafast Dynamics of the Rydberg States of CO$_{2}$: Autoionization and Dissociation Lifetimes}

\author{D. Biswas$^1$, J. K. Wood$^2$, I. Shalaby$^1$, A. Sandhu$^{*1,2}$}

\affiliation{$^1$Department of Physics, University of Arizona, Tucson, Arizona 85721, USA}
\affiliation{$^2$College of Optical Sciences, University of Arizona, Tucson, Arizona 85721, USA}
\date{\today}
\email{asandhu@arizona.edu}

\begin{abstract}
    We report the measurement of ultrafast relaxation dynamics of excited states of carbon dioxide molecule using time-resolved pump-probe photoelectron spectroscopy. Neutral ground state carbon dioxide is excited to $nd\sigma_g$ Henning sharp Rydberg states with an attosecond extreme ultraviolet pulse train. A time delayed near infrared probe pulse is used to photoionize these states to their corresponding ionization limit $B^2\Sigma_u^+$. We obtain differential kinetic energy spectrograms and angular distributions for photoionization and autoionization channels. We model the competition between predissociation and autoionization in the Rydberg state dynamics and analyze differential photoelectron yield as a function of the time delay to extract previously unknown autoionization and predissociation lifetimes for three Henning sharp states (n = 4, 5, 6). 
\end{abstract}

\maketitle

\section{Introduction}
Ultrafast femtosecond and attosecond light pulses have enabled time-resolved probing of molecular dynamics. In particular, time-resolved photoelectron spectroscopy has served as an important spectroscopic tool to study photoionization and excited state dynamics in complex molecular systems \cite{stolow2004femtosecond, schuurman2022time, blanchet2001electronic}. Specifically, pump-probe time-resolved photoelectron spectroscopy has been used to gain new insights about autoionization and predissociation in Rydberg state dynamics \cite{Plunkett2019Oxygen, fushitani2016single}, non-adiabatic dynamics \cite{karashima2021ultrafast, suzuki2012time, wu2015excited}, and vibronic wavepacket dynamics \cite{zhang2023effects, blanchet1999discerning} in molecules.

Here we report an experimental study of the Rydberg state dynamics of carbon dioxide (CO$_2$) molecule. It is a linear triatomic molecule with several Rydberg series of interest. Ground state electronic configuration of CO$_2$ is given by $[core](4\sigma_g)^2(3\sigma_u)^2(1\pi_u)^4(1\pi_g)^4$ with symmetry $X^1\Sigma_g^+$ \cite{shaw1995study}. The lowest three ionic continua $(1\pi_g)^{-1}X^2\Pi_g$, $(1\pi_u)^{-1}A^2\Pi_u$ and $(3\sigma_u)^{-1}B^2\Sigma_u^+$ are at 13.778, 17.314 and 18.077 eV adiabatic ionization energies, respectively \cite{shaw1995study, wannberg1984high, liu2000pulsed, mrozowski1942emission, cvejanovic1985threshold}. The Rydberg series associated with the $A^2\Pi_u$ continuum is the Tanaka-Ogawa series \cite{tanaka1962rydberg}. The exact assignments of these series is still not very well-known. The most acceptable assignments of these states are given as $(1\pi_u)^{-1}nd\delta_g$ \cite{tanaka1962rydberg}. Tanaka-Ogawa Rydberg series also has seven known vibrational progressions. All these vibrational levels converge to corresponding vibrational levels of $A^2\Pi_u$ continuum \cite{tanaka1962rydberg, parr1994selective, baltzer1996study}. The $B^2\Sigma_u^+$ continuum has two Rydberg series associated with it, namely Henning sharp $(3\sigma_u)^{-1}nd\sigma_g$ and Henning diffuse $(3\sigma_u)^{-1}ns\sigma_g$ series \cite{henning1932absorptionsspektren, tanaka1960higher, tanaka1962rydberg}. 

A number of photoabsorption, photoionization, and photoelectron studies have been used to study the dynamics of these Rydberg states. Between 17-18 eV excitation energy, the photoexcitation spectrum is dominated by Henning sharp and diffuse series. High resolution synchrotron radiation studies have precisely measured the photoexcitation and photoionization cross-sections of these states \cite{cook1966absorption, shaw1995study, berg1994synchrotron, huestis2011critical}. Recent ultrafast time-resolved studies have also measured the asymmetry parameters in photoelectron distribution with branching ratios between different vibrational states \cite{furch2013photoelectron}. These Rydberg states can undergo spontaneous preferential autoionization to the $X^2\Pi_g$ continuum or they can predissociate to neutral fragments. Autoionization is an electronic correlation driven process whereas predissociation dynamics involves both nuclear and electronic motions. It is well-known from theoretical quantum efficiency calculations as well as fluorescence emission spectroscopy that once these Rydberg states are excited there is a competition between autoionization and neutral predissociation \cite{ukai1992autoionizing}. Despite the extensive interest, there is very little quantitative information available on the dissociation and autoionization lifetimes of these Rydberg states. In a recent study by Fidler et al \cite{fidler2022state} autoionization lifetimes of a few Henning diffuse [$(3\sigma_u)^{-1}ns\sigma_g$] states have been measured for the first time using four-wave-mixing spectroscopy and compared with theoretical calculations from Fano lineshapes. Another recent theoretical study by Pranjal et al \cite{pranjal2023resonant} calculates the valence shell resonant photoionization and molecular frame photoelectron angular distributions for these states. In our experiment, we use photoelectron spectroscopy to report previously unknown predissociation and autoionization lifetimes for three Henning sharp states [$(3\sigma_u)^{-1}nd\sigma_g$, n = 4, 5, 6] to give the first experimental measurement of the relaxation dynamics of these states.

\begin{figure}
\includegraphics[width=\columnwidth]{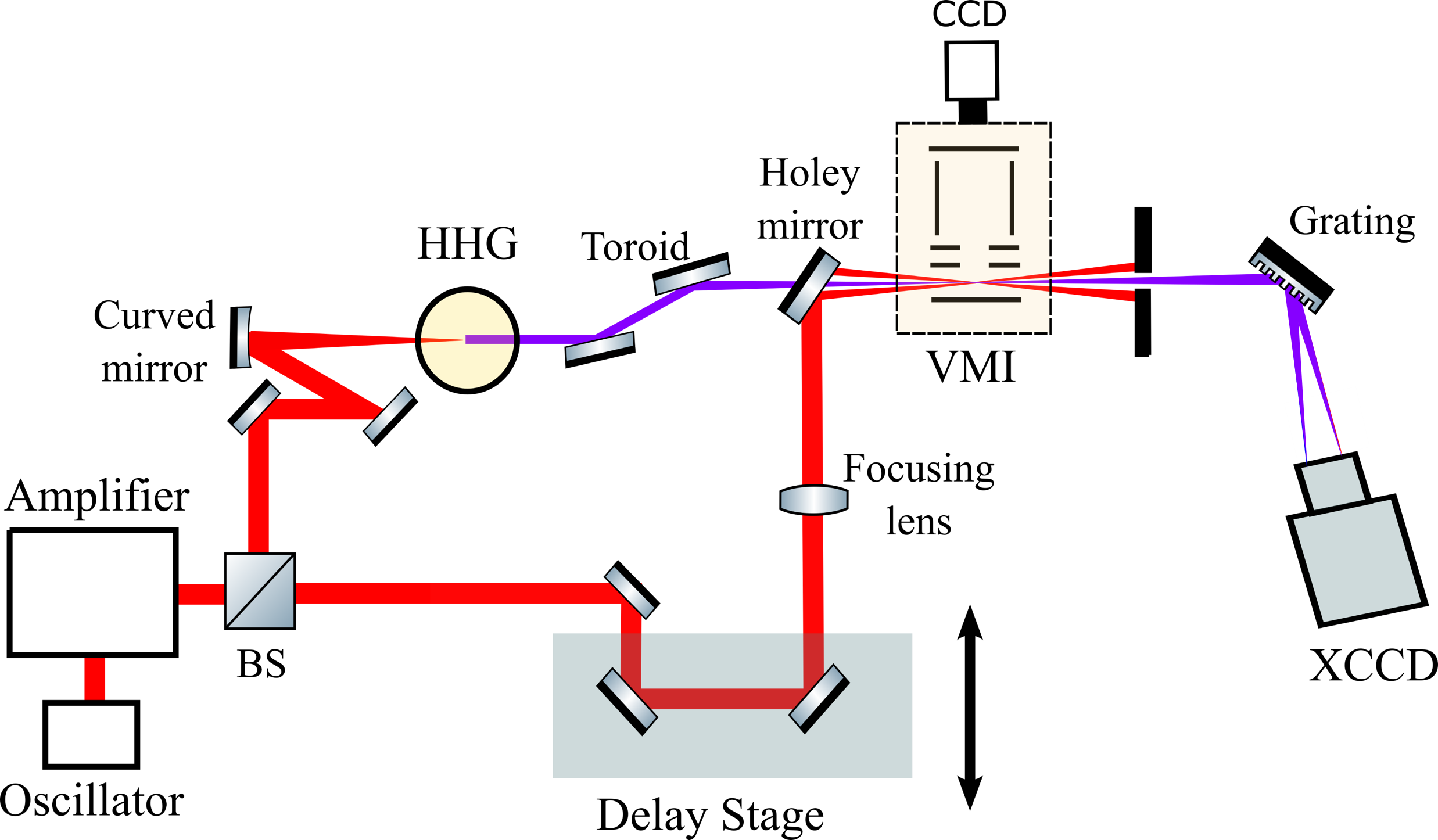}
\caption{Schematic of our experimental apparatus showing a $\sim50$~fs, $1.5$~mJ linearly polarized near infrared (NIR) beam with central wavelength $780$ nm split into pump and probe arms. One arm is focused into a Xe-filled HHG gas cell generating an attosecond XUV pulse train (pump). The other arm (probe) is time delayed and combined with the pump beam on a flat mirror with a hole. Both pump and probe are focused onto an effusive gas jet of carbon dioxide inside of our VMI spectrometer.}
\label{fig:Setup}
\end{figure}

\section{Experiment}

We use a pump-probe experimental setup (Fig. \ref{fig:Setup}) for our photoelectron spectroscopy. We start with a titanium-sapphire laser amplifier system to generate a $\sim 50$ fs near infrared (NIR) Gaussian pulse with a central wavelength of 780 nm, pulse energy of $\sim 1.5$ mJ and repetition rate of 1 kHz. This pulse is split into two equal parts using a 50-50 beamsplitter. One half of the beam is focused into a semi-infinite gas cell filled with xenon to drive high harmonic generation (HHG) and produce attosecond extreme ultraviolet (XUV) pulse trains. This XUV pulse train is composed of only odd harmonics of the driving NIR frequency and used as the pump in our experiment.  The XUV spectral content is separately monitored by using a spectrometer consisting of a grating and an x-ray CCD. Using a toroidal mirror, XUV beam is focused onto an effusive gas jet of carbon dioxide in a velocity map imaging (VMI) spectrometer. The other half of the laser pulse is delayed with respect to the pump and focused using a 50 cm lens to the same interaction point in the VMI spectrometer. The intensity of the probe pulse at the interaction region is $\sim 1$ TW/cm$^2$. A flat mirror with a hole in the center is used to mix the pump and the probe collinearly. Both XUV and NIR pulses are linearly polarized parallel to each other.

The VMI approach \cite{eppink1997velocity} is a well-known and effective approach in the field of photoelectron spectroscopy. In this technique, the three-dimensional momentum distribution of the photoelectrons emanating from an extended focal volume is mapped onto a two-dimensional position-sensitive detector, thus yielding detailed information about both energy and angular distribution of the photoelectrons. The VMI spectrometer has an integrated effusive gas jet of 70 $\mu$m diameter at the center of the repeller electrode. The photoelectron momentum images are recorded using a CCD camera that images the phosphor screen placed behind the microchannel plates (MCPs). The photoelectron  images at each pump-probe time delay are Abel inverted using polar basis function expansion (pBASEX) algorithm \cite{garcia2004two} to obtain 3D photoelectron momentum distribution.  Angular integration of momentum distribution can be used to plot the kinetic energy distribution at each pump-probe delay to obtain time resolved spectrogram. The raw VMI images have some asymmetry due to the non-uniform spatial response of our detector. To address this, we center and symmetrize each VMI image before reconstruction. VMI spectrometer is calibrated using photoelectrons generated from the XUV photoionization of argon. For low energy electrons, small energy shifts were corrected using the photoionization lines from the Rydberg states of Xenon.

We subtract XUV+NIR spectrogram from XUV-only spectrogram at each time delay to extract the delay dependent effect of the probe. In the results discussed below we use this differential photoelectron spectrograms to isolate probe-induced photoionization arising from the XUV excited Rydberg states of CO$_2$, while also analyzing the probe induced disruption of autoionization signals.

\begin{figure}
    \centering
    \includegraphics[width = \columnwidth]{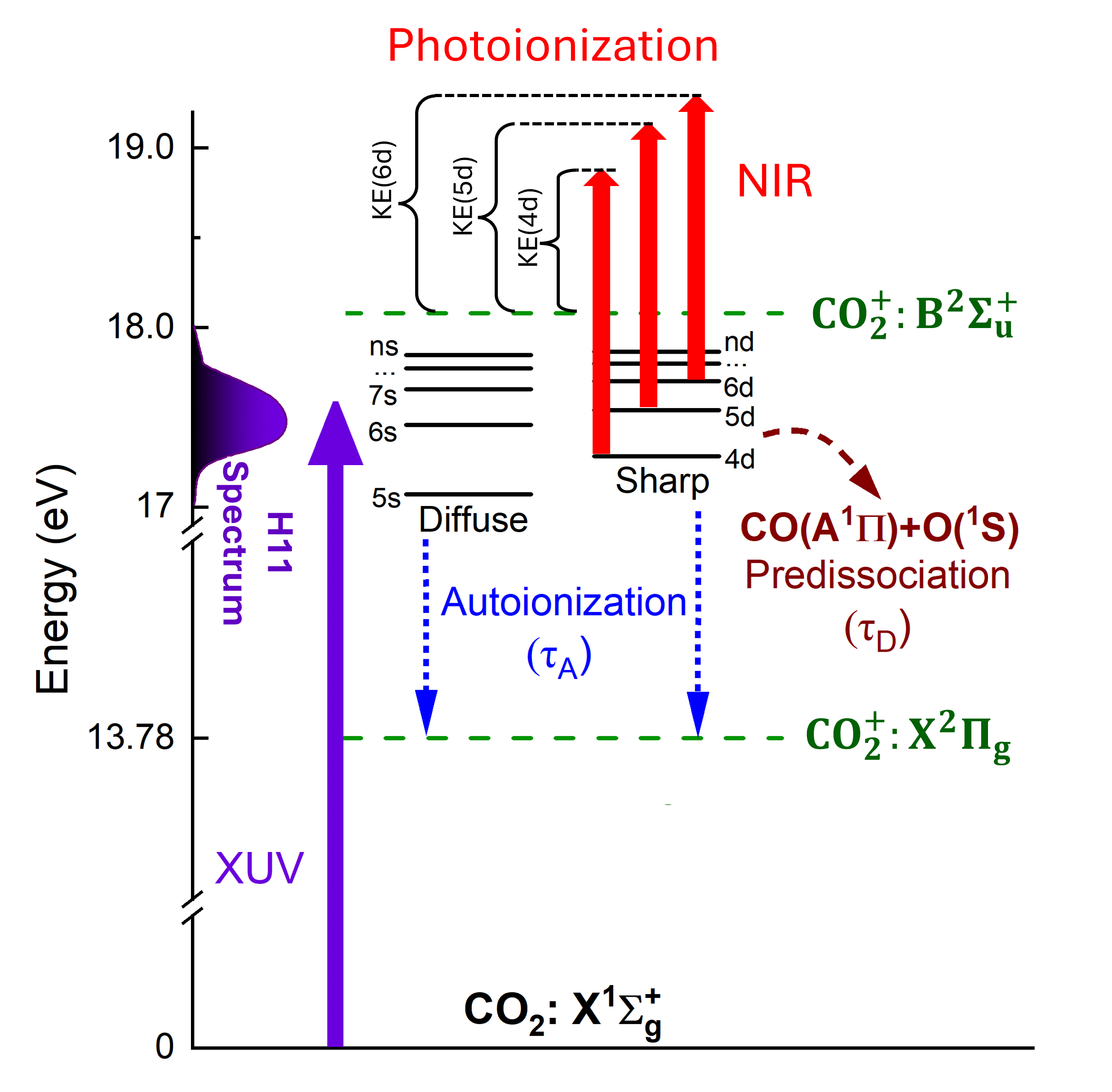}
    \caption{Experimental approach showing excitation of the ground electronic state of CO$_2$ to Henning sharp $(nd\sigma_g)$ and Henning diffuse $(ns\sigma_g)$ Rydberg series with $11^{th}$ harmonic XUV pump. Both these series converge to $B^2\Sigma_u^+$ ionic continuum. Once excited, these Rydberg states can undergo autoionization to $X^2\Pi_g$ state (dotted blue arrows) or predissociate to neutral fragments CO$(A^1\Pi)$ + O$(^1S)$ (dashed brown arrow). Respective timescales associated with these processes $\tau_A$ and $\tau_D$ are shown. An NIR probe pulse at time delay $\tau$ photoionizes the Rydberg states causing depletion of both autoionization and predissociation channels. Expected photoelectron kinetic energies for NIR photoionization are shown for three Henning sharp states (4d, 5d, and 6d) observed in our experiment.}
    \label{fig:schematic}
\end{figure}

The schematic of the XUV excitation and NIR probing in our experiment is shown in the Fig. \ref{fig:schematic}. The 11$^{th}$ harmonic in the XUV spectrum resonantly excites neutral carbon dioxide molecule from the ground state ($X^1\Sigma_g^+$) to a number of neutral Rydberg states between 17-18 eV. The spectrum of the 11$^{th}$ harmonic is peaked around 17.5 eV with a FWHM bandwidth of $\sim$0.4 eV. As mentioned before, there are three different Rydberg progressions of CO$_2$ that has been reported at this energy range: Henning sharp series [$(3\sigma_u)^{-1}nd\sigma_g$], Henning diffuse series [$(3\sigma_u)^{-1}ns\sigma_g$] and, higher vibrational states of Tanaka-Ogawa series [$(1\pi_u)^{-1}nd\delta_g$)]. Both Henning sharp ($nd\sigma_g$) and diffuse ($ns\sigma_g$) series converge to $B^2\Sigma_u^+$ ionic continuum at 18.08 eV. Only these series are shown in Fig. \ref{fig:schematic} because they have significantly larger photoabsorption cross-section than the higher vibrational states of Tanaka-Ogawa progressions. Once excited, the Rydberg states can undergo spontaneous autoionization to $X^2\Pi_g$ ionic continuum at 13.78 eV, or they can predissociate to neutral fragments $CO(A^1\Pi)+O(^1S)$. In general, autoionization and predissociation decay pathways compete, and the associated timescales ($\tau_A$ and $\tau_D$ respectively) have not been fully explored for these states. Measurement of photoionization quantum efficiency indicates that predissociation is dominant in sharp series only \cite{tanaka1962rydberg, shaw1995study}.  We use a probe pulse with photon energy of $\sim1.59$ eV at time delay $\tau$, to photoionize both the molecular Rydberg states as well as the neutral dissociation fragments available at that time delay. Thus, probe induced photoionization restricts the possibility of spontaneous autoionization and we isolate the effect of the probe pulse using differential photoelectron signal as outlined above. We observe that photoionization and autoionization photoelectron angular distributions (PADs) are mutually perpendicular. This  arises from the preferential autoionization direction involving change in core ion symmetry, and it allows us to identify the symmetry of participating Rydberg states. We model the photoelectron yield due to the competition between autoionization and predissociation to estimate lifetimes ($\tau_A$ and $\tau_D$) associated with two decay pathways. Finally, by fitting the time-resolved differential spectrograms we obtain the previously-unknown autoionization and predissociation rates for several members of the Henning sharp Rydberg series of CO$_2$.

\section{Results}
\subsection{Differential Spectrogram}
\begin{figure}
\centering
\includegraphics[width=\columnwidth]{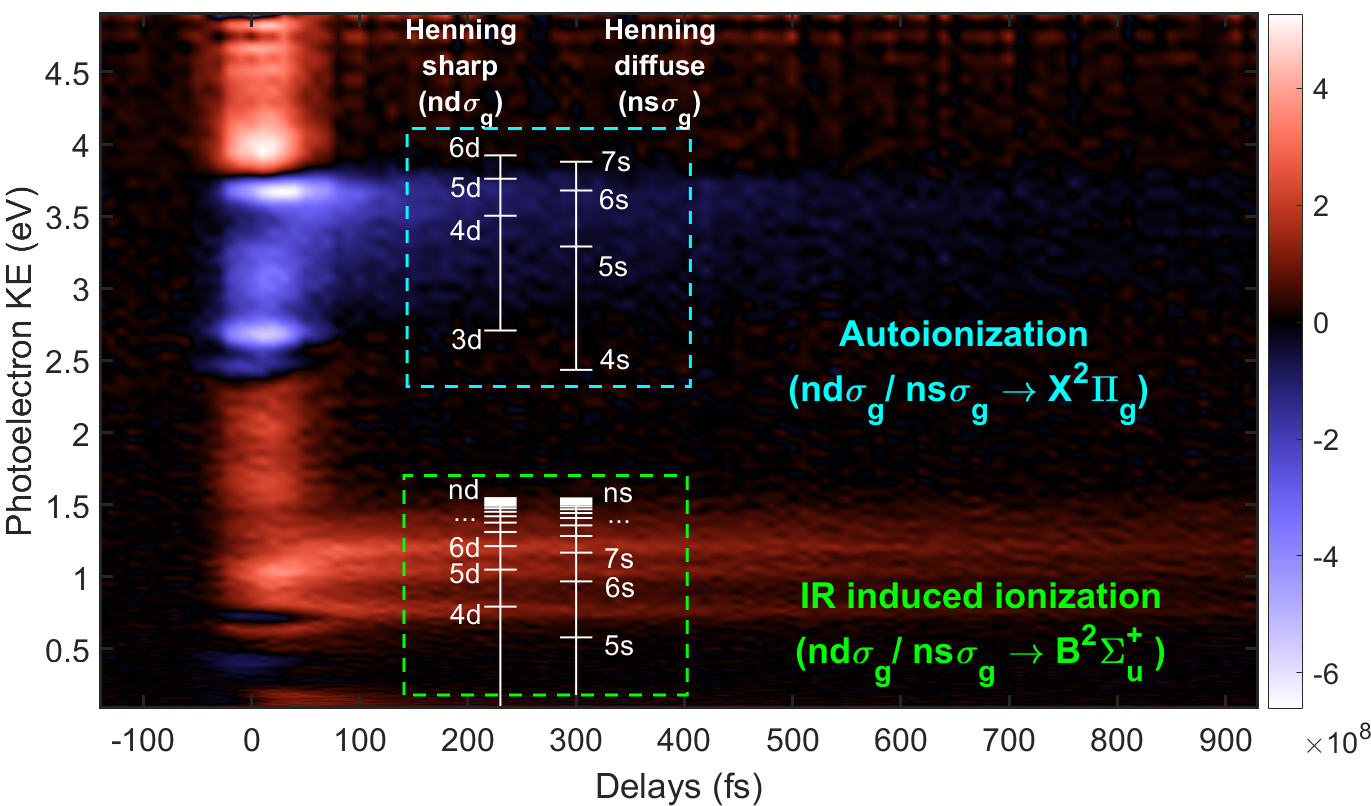}
\caption{Experimentally observed angle integrated time resolved differential photoelectron spectrum of the Rydberg states of CO$_2$. Red and blue implies the NIR probe induced enhancement or depletion of photoelectron yield. Near time delay zero we see direct electrons from XUV-NIR cross-correlation. At positive time delay we see long lived Rydberg states. Expected photoelectron kinetic energies for both autoionization and single photon photoionization are shown for Henning sharp ($nd\sigma_g$) and Henning difuse ($ns\sigma_g$) states.}
\label{fig:spectrogram}
\end{figure}
In Fig. \ref{fig:spectrogram}, the time resolved differential photoelectron spectrogram is shown. At each time delay, momentum images are recorded, reconstructed, and angularly integrated, to obtain kinetic energy as a function of time delay. To isolate the effect of the probe we record both (XUV+NIR) and XUV only spectra at each time delay and subtract them to get a differential spectrum. The spectrogram in Fig. \ref{fig:spectrogram} shows the time evolution of continuum photoelectron kinetic energies at pump-probe time delay interval of 10 fs over a delay range of -150 fs -- 950 fs. Red and blue color scales correspond to positive and negative differences between (XUV+NIR) and XUV-only spectrograms, respectively. Thus, red is the enhancement and blue is the depletion in the photoelectron yield due to the effect of time delayed probe pulse. At negative time delay, the NIR probe precedes the XUV pump excitation, showing no dynamics. Around time delay zero, we see single NIR-photon scattering sidebands of several direct photoionization bands of the neutral CO$_2$ which correspond to emergence of electrons in different ionization continua under the action of the XUV pulse train. These photoelectron bands around time delay zero correspond to XUV-NIR cross-correlation and gives information about the temporal profile of the NIR pulse.

In the positive time delay side of the spectrogram, we see long lived red and blue signals that correspond to the Rydberg state dynamics of the system. Once the 11$^{th}$ harmonic XUV pump excites neutral CO$_2$ to Henning sharp and diffuse series, these are further ionized to $B^2\Sigma_u^+$ by the delayed NIR probe pulse. The long-lived red signals between 0.6-1.5 eV kinetic energies correspond to photoelectron yield due to NIR photoionization. Correspondingly, the long-lived blue negative signals between 3-4 eV kinetic energies represent the depletion of autoionization decay channel by the action of the same NIR pulse. The differential signal analysis allows us to clearly isolate the time-delayed Rydberg state dynamics by subtracting out the contribution of direct XUV photoionization. Expected kinetic energies for both photoionization and autoionization signals for different Rydberg states are shown with white tick marks and energy ladders in Fig. \ref{fig:spectrogram} for both Henning sharp ($nd\sigma_g$) and diffuse ($ns\sigma_g$) series. Blue autoionization signal does not show any discrete energy structure due to the fact that these Rydberg states can autoionize to multiple vibrational levels of $X^2\Pi_g$ ionic continuum thus producing different kinetic energy electrons from a single Rydberg state. On the other hand, low energy red photoionization signal shows three discrete long-lived lines. 

\begin{figure}
\centering
\includegraphics[width=\columnwidth]{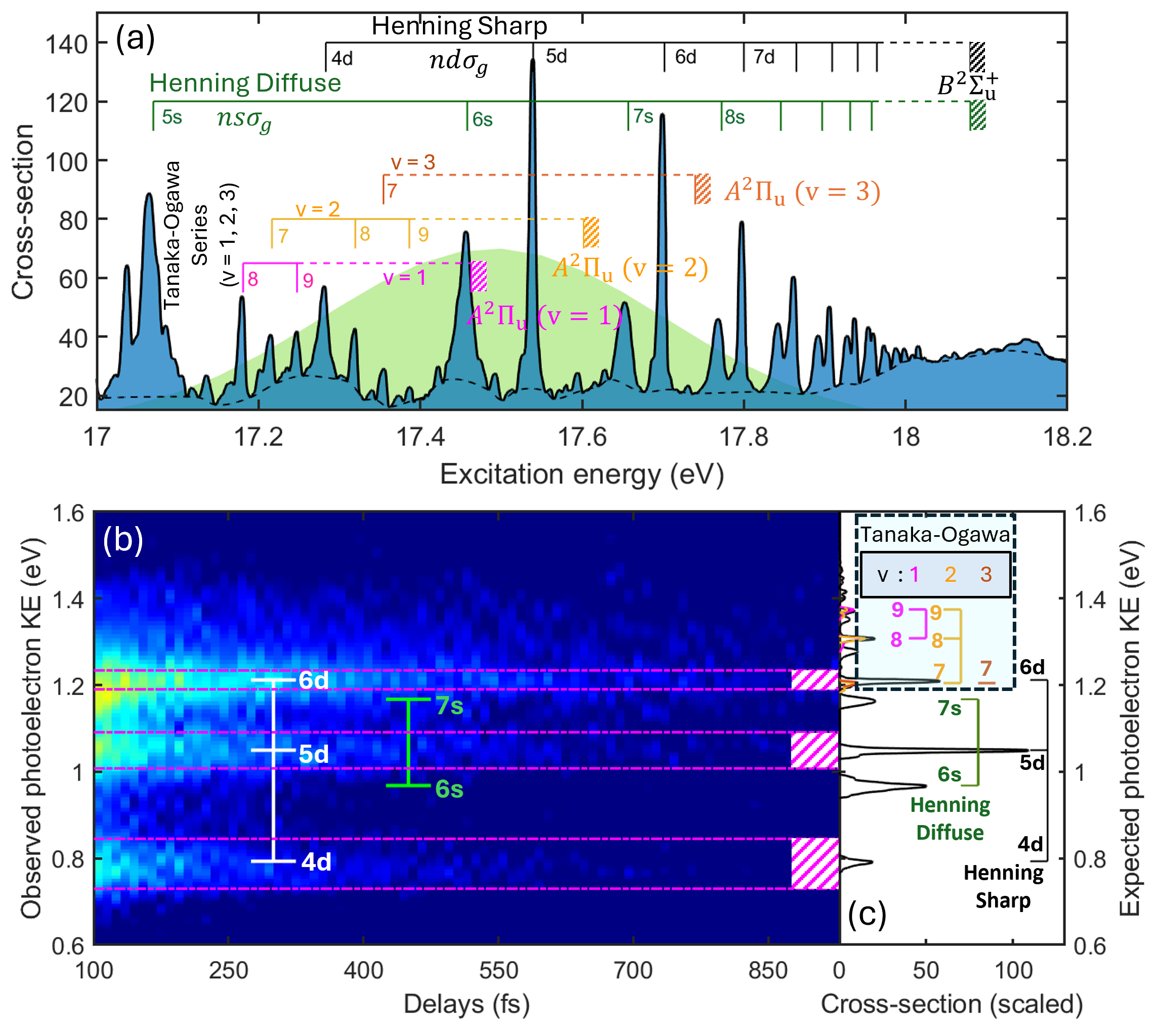}
\caption{(a) Photoabsorption cross-sections of neutral Rydberg states of CO$_2$ in the energy range 17-18 eV (adapted from \cite{huestis2011critical}). Henning sharp ($nd\sigma_g$) and Henning diffuse ($ns\sigma_g$) series are shown with black and green respectively, with their convergence limit ($B^2\Sigma_u^+$) at 18.08 eV. Three Tanaka-Ogawa progressions are also shown for vibrational levels $v = 1$ (magenta), $v = 2$ (yellow) and $v = 3$ (orange) with corresponding limits ($A^2\Pi_u$, $v = 1,2,3$). A typical spectrum for $11^{th}$ harmonic XUV pump is shown with the green shaded area in the background. (b) Differential spectrogram for low energy photoionization signals at positive time delay ($\tau>100$ fs). Expected kinetic energies for single photon ionization of Henning sharp and diffuse states are superposed on the date. Energy windows used for extraction of lifetime and angular distribution for these states are shown with magenta lines and the bound area in between (shown with shaded rectangles). (c) Photoexcitation yield is scaled according to our broadband XUV spectrum and the Rydberg states from panel (a) are then translated to the expected continuum photoelectron energies. Photoelectron energies are calculated separately for each series by considering single NIR photon ionization of the states to the corresponding ionization limits shown in (a).
}
\label{fig:integrated_spec}
\end{figure}

To identify the Rydberg states responsible for these photoionization lines, we use photoabsorption data from the literature \cite{huestis2011critical} as shown in Fig. \ref{fig:integrated_spec}(a). A typical XUV spectrum of the $11^{th}$ harmonic is also shown. We observe that Henning sharp and diffuse series  dominate the spectrum beyond $17.4$ eV. Below that energy, some of the Tanaka-Ogawa states with v = 1, 2 and 3 vibrational levels are visible with cross-sections comparable to the 4d Henning sharp state. All these series converge to their corresponding ionization limits, which are also listed in Fig. \ref{fig:integrated_spec}(a). To compare this photoabsorption data with our work, we do a baseline subtraction followed by multiplication with our XUV spectral profile, to get the scaled photoexcitation spectra relevant to our XUV pump. For each resonance in this spectrum, we obtain the expected photoelectron kinetic energy by adding the energy of single NIR photon and then subtracting the energy of its corresponding ionization limit. The expected photoelectron spectrum obtained in this fashion is shown in Fig. \ref{fig:integrated_spec}(c). For comparison, our delay-dependent photoelectron spectrogram for positive time delays ($\tau>100$ fs) is shown in Fig. \ref{fig:integrated_spec}(b). Comparing the expected and observed photoelectron kinetic energies between Fig. \ref{fig:integrated_spec}(b) and (c) it is clear that the three long-lived lines in the spectrogram correspond to photoionization of 4d, 5d, and 6d states from the Henning sharp series. In fact, the 4d and 5d lines are clearly isolated from the others. Photoelectrons from $n = 7$ Tanaka-Ogawa states with vibrational levels v = 2 and 3 could energetically overlap with the 6d line, but as seen from the Fig. \ref{fig:integrated_spec}(c), the cross-section for 6d photoexcitation is significantly higher than the Tanaka-Ogawa state. Note that we do not see any Henning diffuse state photoelectron peaks in the spectrogram beyond 100 fs time delay because of their short lifetimes (less than 100 fs) \cite{fidler2022state}. We do not analyze the signals for time delays shorted than 100 fs, as the data gets contaminated with NIR modification of the direct XUV photoelectron bands, as discussed before. The observed 4d, 5d, and 6d Henning sharp lines match the expected positions. To extract the lifetime of the Henning sharp states, we use narrow energy windows around the sharp series as shown in Fig. \ref{fig:integrated_spec}(b) using magenta lines and shaded areas bound by those. 


\subsection{Angular distribution}
\begin{figure*}
\includegraphics[width=0.8\linewidth]{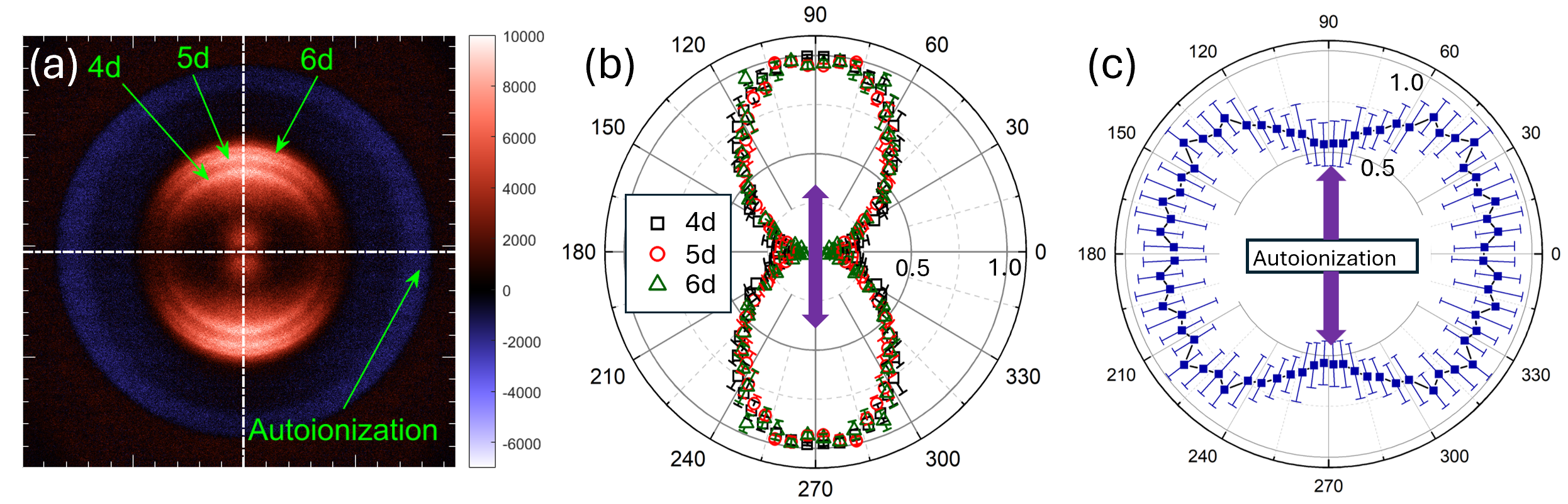}
\caption{Angular dependence of differential photoionization and autoionization signal extracted from centered and symmetrized VMI images. (a) Differential photoelectron VMI image for positive time delay showing three red photoelectron rings for $nd$ Henning sharp states and autoionization electrons in blue, (b) Angular distribution of 4d, 5d, and 6d photoionization electrons at 250 fs time delay.(b) Angular distribution of autoionization depletion signals integrated from 100-250 fs  In each figure, the vertical purple arrow corresponds to the polarization direction for both XUV and NIR pulses.}
\label{fig:angular}
\end{figure*}

The fact that the photoelectron signals correspond to the Henning sharp series is further corroborated by the angular distributions of the differential photoionization and autoionization signals. In Fig. \ref{fig:angular}(a) we show reconstructed VMI image of the differential change in the angular distribution of the electrons integrated over positive time delays from 100 -- 500 fs. As before, the red color represents enhancement due to the NIR probe induced photoionization of 4d, 5d, and 6d Rydberg states. These signals appear as distinct rings in the VMI image as the photoelectrons emerge along the polarization direction of the XUV pump and NIR probe, which are parallel to each other. On the other hand, the blue color represents disruption of the autoionization pathways, leading to depletion in the yield of autoionization electrons. It is clear that the autoionization depletion signal peaks perpendicular to XUV-NIR polarization direction.

Fig. \ref{fig:angular}(b) shows photoelectron angular distributions for 4d, 5d, and 6d states, respectively, where each figure contains three sets of data at  time delay of 250 fs. We find that all three states show nearly identical angular distribution which confirms that all signals are arising from states with same symmetry. We also observed that there is no significant time-dependent change associated with the angular distribution of 4d, 5d, and 6d states within the 100-500 fs timescale. Fig. \ref{fig:angular}(c) shows the angular distribution plot for autoionization depletion integrated from 100-250 fs. Autoionization signal is weaker and more diffuse due to contribution from different vibrational states and it also decays much faster than the photoionization signals, leading to larger error bars.  

The parallel (perpendicular) angular distribution of photoelectrons (autoionizing electrons) elucidates the nature of states excited by the XUV pump. Essentially, the Henning series states are populated due to parallel excitation by the XUV pump pulse and they are further ionized to $B^2\Sigma_u^+$ ionic continuum using a NIR pulse with the same polarization. Thus we expect a dipolar pattern of photoelectron emission along the XUV-NIR polarization direction as seen in the angular distributions. These observations also rule out Tanaka-Ogawa series as the source, which requires a perpendicular excitation from the ground state. The autoionization angular distributions further support these conclusions. When autoionization occurs from $\sigma_g$ symmetry Henning states to $X^2\Pi_g$ ionic continuum, the electrons emerge with one unit of angular momentum. As the electron decays into the $X^2\Pi_g$ states of a molecular ion preferentially aligned along the XUV polarization, the autoionizing electron comes out perpendicular to the light polarization direction.

Thus, using both angular distributions and photoexcitation cross-section data we confirm that the observed photoelectron signals arise predominantly from the Henning series. Furthermore, beyond 100 fs time delay, the contributions are solely from the sharp $nd \sigma_g$ states which have long lifetimes. Referring back to Fig. \ref{fig:integrated_spec}(b), we see that the photoelectron yields from these nd states decay as a function of time delay between pump and probe. Next, we model the competition between autoionization and predissociation decay to extract both lifetimes for the three Henning sharp states observed here.

\subsection{Modeling predissociation and autoionization}
To understand the role of autoionization and predissociation dynamics, we used a model \cite{Plunkett2019Oxygen} for time-delay dependent photoelectron yield taking both decay channels into account. Our XUV pulse excites neutral ground state population to an excited molecular Rydberg state. This population can decay via autoionization with a rate of $\gamma_A$ and via predissociation with a rate of $\gamma_D$. If initial excited state population is $P_0$, time-dependent population in the molecular Rydberg state can be written as
\begin{equation}
    P_{mol}(t) = P_0 e^{-\gamma_At}e^{-\gamma_Dt}.
\end{equation}
The population going into autoionization or predissociation channels in time t can be written as
\begin{equation}
    P_{auto}(t) = P_0(1-e^{-\gamma_Tt})\left(\frac{\gamma_A}{\gamma_T}\right) ,
\end{equation}
\begin{equation}
    P_{diss}(t) = P_0(1-e^{-\gamma_Tt})\left(\frac{\gamma_D}{\gamma_T}\right),
\end{equation}
where $\gamma_T = \gamma_A+\gamma_D$ is the total decay rate of the molecular Rydberg population $P_{mol}$.

Delayed NIR probe can photoionize the population available in either molecular Rydberg state or in predissociated neutral fragments. Both channels will produce photoelectrons of same kinetic energies. Thus, the total population available for NIR photoionization at time t can be written as
\begin{eqnarray}
    P_{tot}(t) && = P_{mol}(t)+P_{diss}(t)\nonumber\\
    && = P_0 e^{-\gamma_Tt}+P_0(1-e^{-\gamma_Tt})\left(\frac{\gamma_D}{\gamma_T}\right)\nonumber\\
    && = P_0 e^{-\gamma_Tt} \left(\frac{\gamma_A}{\gamma_T}\right)+P_0 \left(\frac{\gamma_D}{\gamma_T}\right).
\end{eqnarray}
The total photoionization rate with a Gaussian NIR pulse at time delay $\tau$ is
\begin{equation}
    \frac{dn_e}{dt}(t,\tau) \propto P_{tot}(t) e^{-(t-\tau)^2/2\sigma^2}
\end{equation}
where $\sigma$ is the width of the Gaussian pulse. By integrating over time t we get the photoelectron yield as a function of pump-probe time delay.
\begin{equation}
\label{eqn:photoelectron_yield}
    n_e(\tau) \propto \left(\frac{\gamma_A}{\gamma_T}\right)\mathcal{G}_{exp}(\tau, \gamma_T, \sigma)+\left(\frac{\gamma_D}{\gamma_T}\right)\mathcal{G}_{exp}(\tau, 0, \sigma),
\end{equation}
where $\mathcal{G}_{exp}$ is defined as the exponential convolution of Gaussian function.
\begin{eqnarray}
    \mathcal{G}_{exp}(\tau, \gamma, \sigma) = &&  \int_0^\infty dt e^{-\gamma t}e^{-(t-\tau)^2/2\sigma^2}\nonumber\\
    = &&  \sigma\sqrt{\frac{\pi}{2}}\exp\left(\gamma\tau+\frac{\gamma^2\sigma^2}{2}\right)\nonumber\\
    && \times\text{erfc}\left(\frac{\gamma\sigma}{\sqrt{2}}-\frac{\tau}{\sigma\sqrt{2}}\right)
\end{eqnarray}

\begin{figure}
\includegraphics[width=\columnwidth, height = 1.75\linewidth]{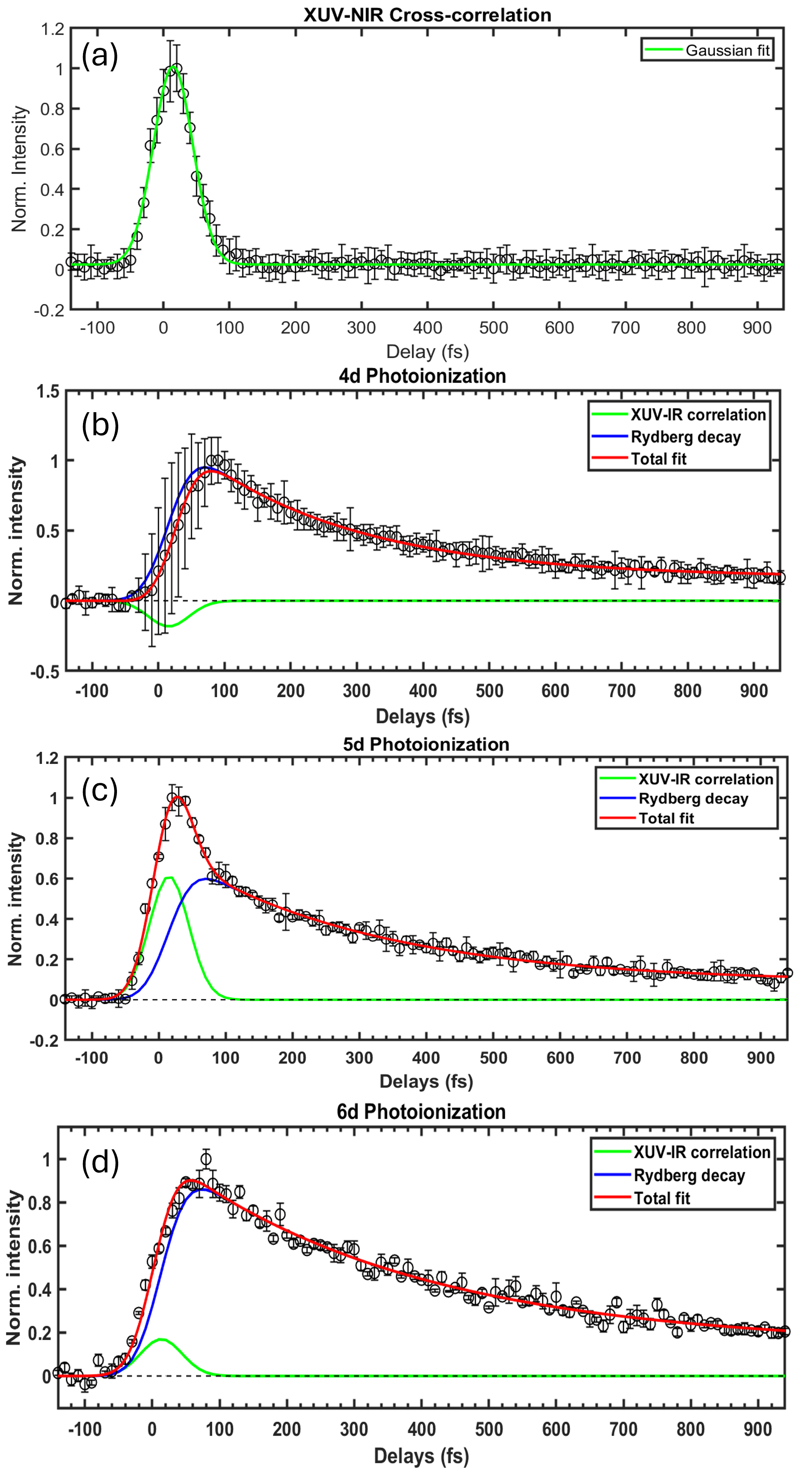}
\caption{Delay dependent lineouts from the differential spectrogram. (a) XUV-NIR Gaussian cross-correlation as measured from the NIR dressing of the direct photoelectrons generated by XUV ionization to $X^2\Pi_g$ continuum by $11^{th}$ harmonic around time delay $\tau\sim0$ fs, (b) NIR photoionization of XUV excited 4d Rydberg state, (c) NIR photoionization of 5d Rydberg state and, (d) NIR photoionization of 6d Rydberg state. In these plots, solid blue lines correspond to photoelectron yields from probe induced photoionization from molecular as well as atomic Rydberg states, solid green lines correspond to contributions from probe induced dressing of direct photoelectrons around time-delay $\sim 0$ fs and, red lines are the sum of both that gives the total fitting.}

\label{fig:lifetimes}
\end{figure}

\subsection{Autoionization and predissociation lifetimes}
Using the model  above we fit experimentally-observed photoelectron yields for 4d, 5d, and 6d Rydberg states to extract autoionization and predissociation lifetimes. In the discussion of the experimental spectrogram Fig. \ref{fig:spectrogram}, we mentioned that there is a contribution to the signal from the pump-probe cross-correlation around time delay zero. To take that into account, we added another Gaussian contribution to the photoelectron yield given in Eq. \ref{eqn:photoelectron_yield}. 

As a first step, we use this Gaussian term to fit the photoelectron yield in the NIR dressing sideband near 1.6 eV (Fig. \ref{fig:lifetimes}(a)). Fitting the XUV-NIR cross-correlation allows us to extract the width of the probe pulse while also calibrating the zero delay. The standard deviation ($\sigma$) width of the Gaussian cross-correlation is obtained to be $29.77$ fs with an estimated error of $3\%$, while the correction for zero time delay between pump and probe pulse is obtained within $\sim 1$ fs accuracy. Using these parameters of the NIR pulse, we fit three Rydberg photoelectron yields as shown in Fig. \ref{fig:lifetimes}(b)-(d). In each plot, green line represents the XUV-IR cross-correlation signal, blue line corresponds to the Rydberg state dynamics, while the red line is a fit to the total signal. Using this method, we extracted the autoionization and predissociation lifetimes as listed below in Table~\ref{tab:table1}.
\begin{table}[h]
\caption{\label{tab:table1}
Autoionization ($\tau_A$) and predissociation ($\tau_D$) lifetimes of 4d, 5d, and 6d Henning sharp series. Numbers in the parenthesis account for the range of fitting parameters for 95\% confidence interval.}

\begin{ruledtabular}
\begin{tabular}{lll}
\multicolumn{1}{c}{\textrm{State}}&
\textrm{$\tau_A$ (fs)}& \textrm{$\tau_D$ (fs)}\\
\colrule
4d & 286 (265, 310) & 1783 (1536, 2124)\\
5d & 339 (310, 374) & 2574 (2003, 3652)\\
6d & 427 (388, 474) & 3055 (2079, 5757)\\
\end{tabular}
\end{ruledtabular}
\end{table}

There is no experimental measurements for the lifetimes of these states available in the literature. Recently, autoionization lifetimes were calculated for these states in Fidler et al \cite{fidler2022state}. Our experimental lifetime for 5d sharp series matches well within the error limits with the calculated value of 372.6 fs, whereas the lifetime of 6d state does not match with the calculated value of 1457.5 fs. We suspect that the complexity of involving the predissociation process with autoionization causes this discrepancy between the calculation and the experiment. 
In the experiment by Fidler et al \cite{fidler2022state}, they observe only Henning diffuse state lifetimes due to two-photon Raman transitions with nearby dark resonances. Apart from the experimental measurments of the autoionization lifetimes of Henning sharp states, we have also obtained the experimental predissociation lifetimes of these states. Although there is a significantly-large error bar on the predissociation lifetimes, there is no calculation or experimental measurement of these lifetimes available in literature.

\section{Conclusion}
In this work we used ultrafast time-resolved photoelectron spectroscopy to study the Rydberg state dynamics of CO$_2$ molecule. Using delay-dependent photoelectron kinetic energy and angular distribution measurements we identified strong signals for three Henning sharp Rydberg states (4d, 5d, and 6d). Using a model that accounts for the competition between autoionization and predissociation, we report the first experimental measurements of both autoionization and predissociation lifetimes for these three Henning sharp states. This XUV-NIR photoelectron spectroscopy method can be generalized to observe the non-adiabatic relaxation dynamics associated with electronic correlations and electron-nuclear couplings in a variety of polyatomic molecules. The knowledge of relaxation rates and dynamics of excited molecules has implications for understanding the charge and energy redistributions in a variety of systems. 


\section{Acknowledgments}
This work was supported by the National Science Foundation under grant numbers PHY 2207641 and PHY 1919486.

\bibliography{Refs}
\end{document}